\documentclass[aps,amsmath,amssymb,showpacs,floatfix,twocolumn,prl,superscriptaddress]{revtex4}
\usepackage{graphicx}
\usepackage{dcolumn}
\usepackage{bm}

\def\vk{{\mathbf k}}
\def\ek{\varepsilon_{{\mathbf k}}}

\begin{document}

\title{Valence-Bond Dynamical Mean-Field Theory of Doped Mott Insulators\\
with Nodal/Antinodal Differentiation}

\date{\today}

\author{Michel Ferrero}
\affiliation{Centre de Physique Th{\'e}orique, CNRS, {\'E}cole
Polytechnique, 91128 Palaiseau Cedex, France.}
\author{Pablo S. Cornaglia}
\affiliation{Centro At{\'o}mico Bariloche and Instituto Balseiro,
CNEA, CONICET, 8400 Bariloche, Argentina.}
\affiliation{Centre de Physique Th{\'e}orique, CNRS, {\'E}cole
Polytechnique, 91128 Palaiseau Cedex, France.}
\author{Lorenzo De Leo}
\affiliation{Centre de Physique Th{\'e}orique, CNRS, {\'E}cole
Polytechnique, 91128 Palaiseau Cedex, France.}
\author{Olivier Parcollet}
\affiliation{Institut de Physique Th{\'e}orique, CEA, IPhT, CNRS, URA 2306, F-91191 Gif-sur-Yvette, France}
\author{Gabriel Kotliar}
\affiliation{Physics Department and Center for Materials Theory, Rutgers University,
Piscataway NJ 08854, USA }
\author{Antoine Georges }
\affiliation{Centre de Physique Th{\'e}orique, CNRS, {\'E}cole
Polytechnique, 91128 Palaiseau Cedex, France.}

\begin{abstract}
We introduce a valence-bond dynamical mean-field theory of
doped Mott insulators. It is based on a minimal cluster of
two orbitals, each associated with a different region of momentum
space and hybridized to a self-consistent bath.
The low-doping regime is characterized by singlet formation and
the suppression of quasiparticles in the antinodal regions, leading to
the formation of Fermi arcs.
This is described in terms of an orbital-selective transition in
reciprocal space.
The calculated tunneling and photoemission spectra are consistent with
the phenomenology of the normal state of cuprates.
We derive a low-energy description of these effects using a
generalization of the slave-boson method.
\end{abstract}

\maketitle

The doping of a Mott insulator is a fundamental problem of condensed matter physics,
relevant to the physics of
cuprate superconductors~\cite{Anderson_1987}.
In the simplest Brinkman-Rice~\cite{Brinkman1970} description, the doped metallic state is a Fermi liquid in which quasiparticles
are formed with a heavy mass $m^*/m\sim1/\delta$ and a
reduced weight $Z\sim\delta$ ($\delta$ is the doping level).
This physical picture can indeed be rationalized using the modern theoretical
framework of dynamical mean-field theory (DMFT)~\cite{georges1996,kotliar_review_rmp_2006}.
It is applicable when spatial correlations are weak, which is favored by
high dimensionality and strong competing (e.g. orbital) fluctuations.
In cuprates however, which are two-dimensional materials with low orbital
degeneracy, it was pointed long ago by Anderson in a seminal paper~\cite{Anderson_1987}  that
the antiferromagnetic superexchange ($J$) plays a key role,
leading to strong short-range correlations associated with singlet formation (valence bonds) between
nearest-neighbor lattice sites.
Slave-boson mean-field theories~\cite{Baskaran_1987,Kotliar_1988,Fukuyama88,LeeRMP06},
as well as projected variational wave-functions~\cite{Gros_1988,Paramekanti_2001},
provide simple theoretical frameworks to
incorporate this effect, modifying the Brinkman-Rice
picture at small doping $\delta\lesssim J/t$ and leading, e.g. to
a finite effective mass $m^*/m\sim 1/(J/t+\delta)$, consistent
with observations in cuprates.
However, these theories fail to describe a key phenomenon
in underdoped cuprates, namely the strong differentiation in momentum space
observed e.g. by photoemission spectroscopy (ARPES)~\cite{damascelli_rmp_2003}:
Coherent quasiparticle excitations are suppressed in the antinodal regions of the Brillouin zone (BZ)
and a pseudogap appears in the normal state.
In order to take this phenomenon into account while incorporating short-range
correlations, cluster extensions of the DMFT framework have been
investigated by several groups~\cite{maier_cluster_rmp_2005,kotliar_review_rmp_2006,Tremblay06}.
Most studies have considered clusters of at least four sites (plaquette) and
numerical efforts have been devoted to increase the cluster size in order to improve
momentum-resolution and get closer to the two-dimensional lattice~\cite{MaierSCLargeCluster}.

In this article, we follow a different route, looking for
a description based on the {\it minimal} cluster
able to successfully describe momentum-space differentiation together with Mott physics.
We find that a two-site cluster is sufficient to achieve this goal on a qualitative level,
and to a wide extent on a quantitative level when compared to larger cluster calculations.
This allows us to construct a valence-bond dynamical mean-field theory (VB-DMFT)
of nodal/antinodal differentiation, in which this phenomenon is linked to the
distinct properties of the orbitals associated with different regions of momentum-space.

The main motivation to choose the smallest possible cluster is
to  advance our qualitative understanding.
Since the theory is based on a two-site Anderson model,
results can be interpreted in terms of
valence-bond singlet formation and linked to the competition
between singlet-formation
and individual Kondo screening~\cite{jones88,FerreroNutshell,deleo2004,Schiro08}.
However, the self-consistency of the bath does bring in novel aspects to this competition
in comparison to the non self-consistent two-impurity model.
The present VB-DMFT can be viewed as an extension of the static mean-field
theories (e.g. slave-boson based) of singlet formation. In contrast to those theories
which have a limited number of static variational parameters, it involves
a dimer coupled to a self-consistent bath through energy-dependent hybridization functions.
This allows for a description of the physics over a wide range of energy scales.
At low energy however, the new slave boson approximation introduced in~\cite{Lechermann2007a}
reproduces with a remarkable accuracy several
aspects of the full solution.

We study the Hubbard model on a square lattice, with hopping between nearest-neighbor
($t$) and next-nearest-neighbor sites ($t'$). In the following, we use
 $U/t=10$ and $t'/t=-0.3$, which are values commonly used for modeling
hole-doped cuprates in a single-band framework.
All energies (and temperatures) are expressed in units of $D=4t=1$, and
the doping is denoted by $\delta$.
We use a two-site effective Anderson impurity problem, involving the on-site
interaction $U$ and two hybridization functions: a local one $\Delta_{11}(\omega)=\Delta_{22}(\omega)$
and an inter-site one $\Delta_{12}(\omega)$, which are self-consistently
determined by relating the two-impurity problem to the original lattice one. We have investigated
several such embeddings, both of the dynamical cluster approximation (DCA)
and cellular-DMFT (CDMFT) type~\cite{maier_cluster_rmp_2005,kotliar_review_rmp_2006}
with similar results. Here, we focus on a somewhat generalized form of the DCA embedding, which
preserves the symmetries of the square lattice, in which the Brillouin zone
is decomposed into two patches of equal surface: a central square (denoted $P_+$) centered at momentum
$(0,0)$ and the complementary region ($P_{-}$) extending to the edge of the BZ and containing
in particular the $(\pi,\pi)$ momentum.
From the lattice Green's function, two coarse-grained Green's functions in momentum space are
constructed:
$G_{\pm}(\omega)=\sum_{\vk\in P_{\pm}} G(\vk,\omega)$
(with momentum summations normalized to unity within each patch).
Following the
DCA construction, the inner (resp. outer) patch self-energy is associated with the even- (resp. odd) parity
self-energy of the two-impurity effective problem, i.e to the even (resp. odd) orbital combinations
$(c^\dagger_1\pm c^\dagger_2)/\sqrt{2}$.
Indeed, the states close to $(0,0)$ have more bonding character while those close to $(\pi,\pi)$ have
more antibonding character.
The self-consistency condition reads:
$G_{K}(\omega)=\sum_{\vk\in P_{K}} [\omega+\mu - \ek - \Sigma_K(\omega) ]^{-1}$.
In this expression, the index $K=\pm$ refers both to the inner/outer patch index and
to the even/odd orbital combinations.
We solve the self-consistent two-impurity problem using
both continuous-time quantum Monte Carlo (CTQMC)~\cite{CTQMC_Werner}
which sums the perturbation theory in
$\Delta_{ab}(i\omega_n)$ on the Matsubara axis,
and an approximate method geared at low-energy properties:
the rotationally invariant slave-boson
formalism (RISB) presented in~\cite{Lechermann2007a}.
The RISB method introduces slave-boson amplitudes $\phi_{\Gamma n}$,
 a density matrix connecting the eigenstates $|\Gamma\rangle$ of the isolated dimer
 to the quasiparticle Fock states $|n\rangle$,
determined by minimizing (numerically) an energy functional.

\begin{figure}[ht!]
  \includegraphics[width=8.5cm,clip=true]{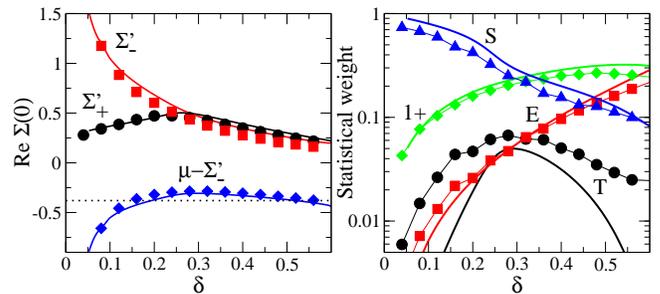}
  \caption{(Color online) Left: real part of $\Sigma_\pm(0)$ as a function of doping level,
  computed with RISB (lines) and
  CTQMC (symbols). $\mu-\Sigma^{\,\prime}_{-}(0)$ (diamonds) reaches the
  odd-orbital band edge (dotted line), which becomes empty at low energy below $\delta\sim 16\%$.
  Right: statistical weights of the various dimer cluster eigenstates.
  $S$ is the intra-dimer singlet, $1+$ the (spin-degenerate) state with
  one electron in the even orbital, $E$ the empty state and $T$ the intra-dimer triplet.
  $\beta = 100$.
  }\label{Fig_ReSig_Weights}
\end{figure}
In Fig.~\ref{Fig_ReSig_Weights}, we display the real part of the even-
and odd-orbital self-energy at zero frequency, as determined by both methods, as
a function of $\delta$. We find a rather remarkable agreement between the
CTQMC solution and the low-energy RISB.
The two orbitals behave in a similar way
at high doping $\delta\gtrsim 25\%$. Below this doping level, we observe an onset of
orbital differentiation, which is a manifestation of momentum differentiation in
the lattice model. This differentiation increases as $\delta$ is reduced, until a transition
is reached at $\delta\simeq 16\%$ (in CTQMC). At this characteristic doping,
$\mu-\Sigma^{\,\prime}_{-}(0)$ reaches
the band edge corresponding to the odd orbital, and the latter becomes empty at low energy
and remains so for all lower dopings.
$G(\vk,\omega)$ no longer has poles
at $\omega=0$ in the outer patch, and low-energy quasiparticles exist
only inside the
inner patch. Hence, at low doping, momentum-space differentiation
becomes strong and manifests itself as an orbital-selective transition in
VB-DMFT.

In order to gain further qualitative insight, we also plot in Fig.~\ref{Fig_ReSig_Weights} (right part)
the statistical weight of several cluster eigenstates $|\Gamma\rangle$,
given within slave bosons
by the amplitude $p_\Gamma=\sum_n |\phi_{\Gamma n}|^2$.
We compare it to a similar estimate~\cite{CTQMC_Haule} from CTQMC.
The agreement between CTQMC and RISB is again
very good, and even quantitative for the two states with highest weights.
At large doping, the empty state and the two spin-degenerate states with one electron in
the even orbital dominate, as expected.
As doping decreases, these states lose weight and the intra-dimer singlet
prevails, reflecting the strong tendency to valence-bond formation.
The states with immediately lower weights are the one-electron states
and the valence-bond breaking triplet excitation which dominates
over the empty state. Therefore, the orbital (momentum) differentiation at low doping
is governed by intra-dimer singlet formation, reminiscent of the singlet regime
of the two-impurity Anderson model.

The gaping of the odd orbital (outer patch)
is actually a crude description of
the pseudogap phenomenon. To illustrate this, we compute the tunneling conductance
$dI/dV$ as a function of voltage $V$. This calculation is made possible
by the high quality, low-noise, of the CTQMC results on the Matsubara
axis, allowing for reliable analytical continuations
to the real axis at low and moderate energy, using simple Pad{\'e} approximants.
\begin{figure}[ht!]
\includegraphics[width=8.0cm,clip=true]{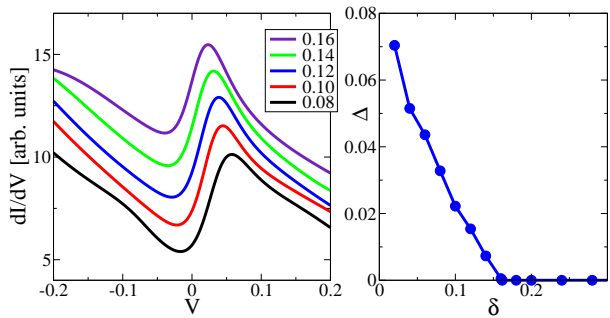}
\caption{\label{Fig.Tunnel}
(Color online) Left panel: tunneling spectra for different doping
levels. All curves are obtained using Eq.~(6) of
Ref.~\cite{ReviewSTMFisherBerthod} with
the same proportionality factor. Right panel:
gap $\Delta$ to the band edge of the unoccupied odd-orbital
states. $\beta =100$.}
\label{Fig_tunneling}
\end{figure}
The conductance is displayed on Fig.~\ref{Fig.Tunnel} together with
the gap $\Delta$ in the odd Green's function, obtained from
$\Delta  = \Sigma^{\,\prime}_{-}(\Delta) + \varepsilon_{\mathrm{min}} - \mu$,
with $\varepsilon_{\mathrm{min}}$ the lower edge of
the band dispersion $\ek$ in the outer patch.
Note the overall particle-hole asymmetry of $dI/dV$ and the peak at positive
voltage. This peak shifts towards higher energy with decreasing doping, as does the gap $\Delta$, and
can indeed be traced back to the edge of the unoccupied odd-orbital spectral function.
These observations compare favorably to tunneling experiments in the normal
state of underdoped cuprates~\cite{ReviewSTMFisherBerthod}.

\begin{figure}[ht!]
  \includegraphics[width=8.5cm,clip=true]{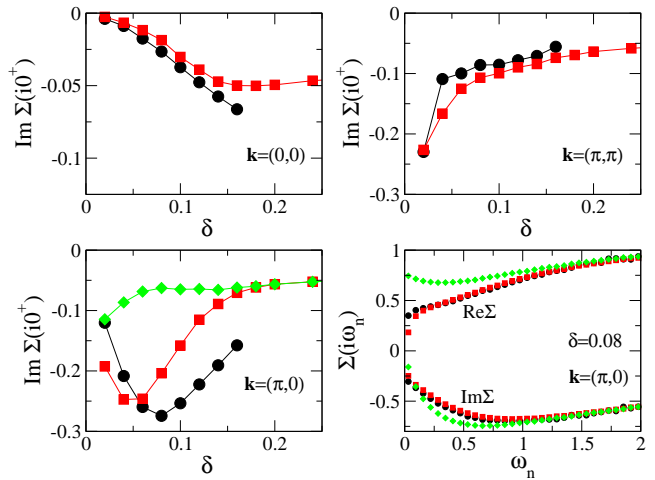}
  \caption{(Color online) Comparison of the plaquette (dots) and dimer self-energies
  at $K = (0,0),(\pi,\pi),(\pi,0)$. Upper and lower-left
  panels: $\mathrm{Im}\Sigma_K(i0^+)$ as a function of doping.
  Lower-right panel: $\Sigma_{\pi 0}(i\omega_n)$ at fixed $\delta=0.08$.
  The dimer results at $K = (\pi,0)$ are obtained by M-interpolation (squares)
  and $\Sigma$-interpolation (diamonds). At $K=(0,0),(\pi,\pi)$ both
  interpolations coincide. $\beta$=100.}
  \label{Fig_link_plaquette}
\end{figure}

We now address two related issues: how to reconstruct information in momentum space
using our two-orbital description, and how to gauge the reliability of a description based
on only two momentum-space components, as compared to calculations with larger cluster
sizes and better momentum-space resolution.
The approximation of lattice quantities from the cluster ones is a
central issue in cluster methods.
Periodization is crucial in CDMFT to restore translational invariance.
In DCA, translational invariance is not broken, but there is still
a large freedom when interpolating the self-energy in the BZ from
the cluster self-energies. The most local quantity is expected to
give the more reliable interpolation. We investigate two possibilities
among those that have been discussed in the literature~\cite{StanescuCumulantCourt}:
{\it i)} interpolating the self-energy ($\Sigma$-interpolation) as
$\Sigma(\vk,\omega)=\Sigma_{+}(\omega)\alpha_+(\vk)+\Sigma_{-}(\omega)\alpha_-(\vk)$,
with $\alpha_\pm(\vk)=\frac{1}{2}\{1\pm\frac{1}{2}[\cos(k_x)+\cos(k_y)]\}$;
{\it ii)} interpolating the cumulant  ($M$-interpolation) as
$M(\vk,\omega)=M_{+}(\omega)\alpha_+(\vk)+M_{-}(\omega)\alpha_-(\vk)$.
The cumulant is related to the self-energy by $\Sigma=\omega+\mu-M^{-1}$.
It is the dual quantity of the self-energy in an expansion
around the atomic limit
and a natural measure of how much the hybridization to the self-consistent environment changes the
impurity Green's function as compared to an isolated dimer.
Close to the Mott insulator, it is more local than the self-energy
and a better quantity to interpolate~\cite{StanescuCumulantCourt}.
Comparing the interpolations on small clusters with
direct calculations on larger clusters, having better $\vk$-resolution,
provides a systematic test of cluster schemes and interpolations.
As a first step, we compare in Fig.~\ref{Fig_link_plaquette}
the results of VB-DMFT, with $\Sigma\,$- or
$M\,$-interpolation, to the cluster components $\Sigma_K(\omega)$
of a four-site cluster (plaquette),
using the standard DCA embedding (with the BZ divided into 4 patches centered around
$K=(0,0), (0,\pi), (\pi,0), (\pi,\pi))$.
The results of Fig.~\ref{Fig_link_plaquette} reveal two main points:
{\it i)} the $M$-interpolation of the
two-orbital results is clearly superior to the $\Sigma$-interpolation for reconstructing plaquette cluster
quantities and
{\it ii)} when $M$-interpolated,
the two-orbital description does quite a remarkable job at capturing the
full frequency-dependence of the various cluster components $\Sigma_K(\omega)$ of the plaquette results.
Note that the plaquette cluster-momentum $K=(\pi,0)$ is not present
as an individual orbital in the two-site description: it is entirely reconstructed
by interpolation, and as such is the most direct test of the reconstructed
momentum-dependence.
A distinctive feature of the results depicted in Fig.~\ref{Fig_link_plaquette} is that the scattering
rate near momentum $(\pi,0)$, $\mathrm{Im}\Sigma_{\pi 0}(i0^+)$,
displays a maximum around a doping level $\delta\simeq 8\%$, as
previously noted in the plaquette study of~\cite{haule_plaquette_2007}.
Note however, that this maximum does not induce a maximum of the scattering rate
computed {\it at the Fermi surface}.

\begin{figure}[ht!]
  \includegraphics[width=4.25cm,clip=true]{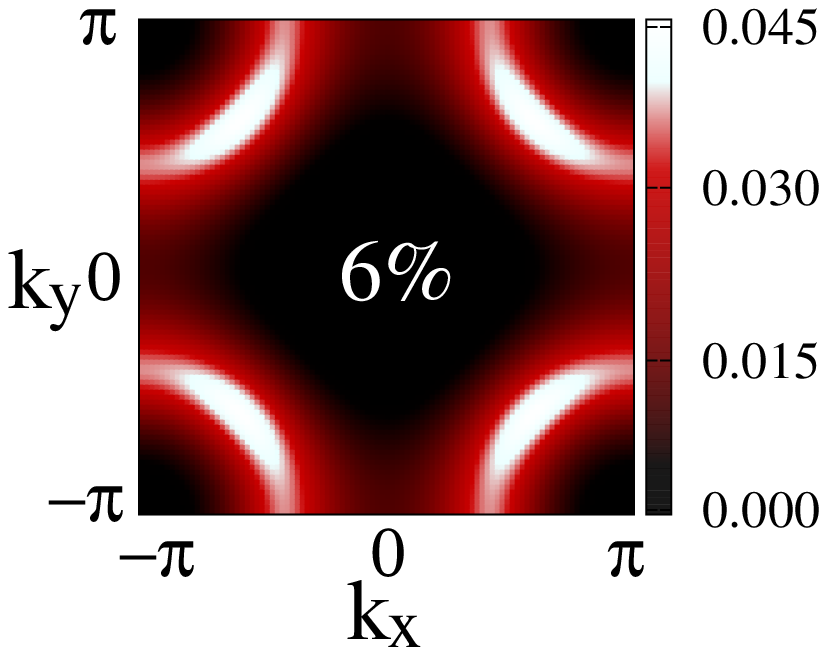}
  \includegraphics[width=4.25cm,clip=true]{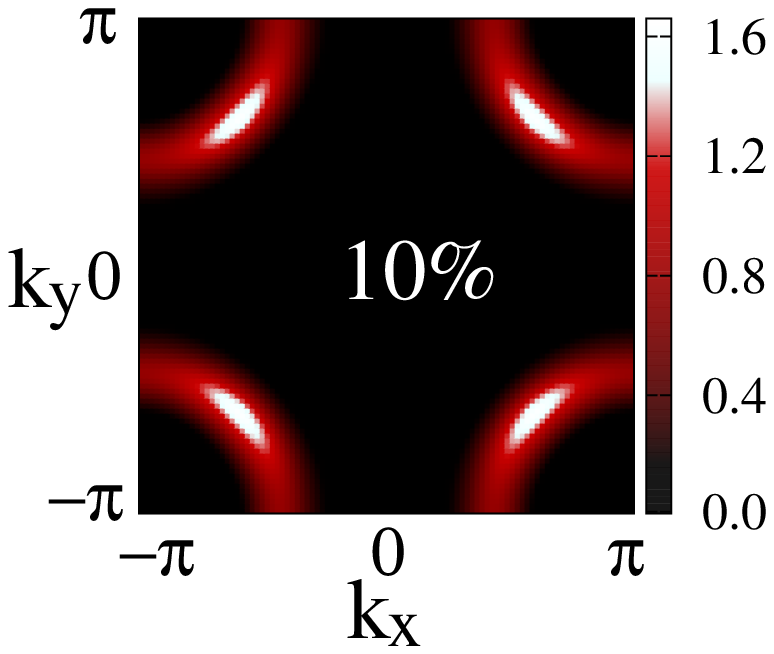}\\
  \includegraphics[width=4.25cm,clip=true]{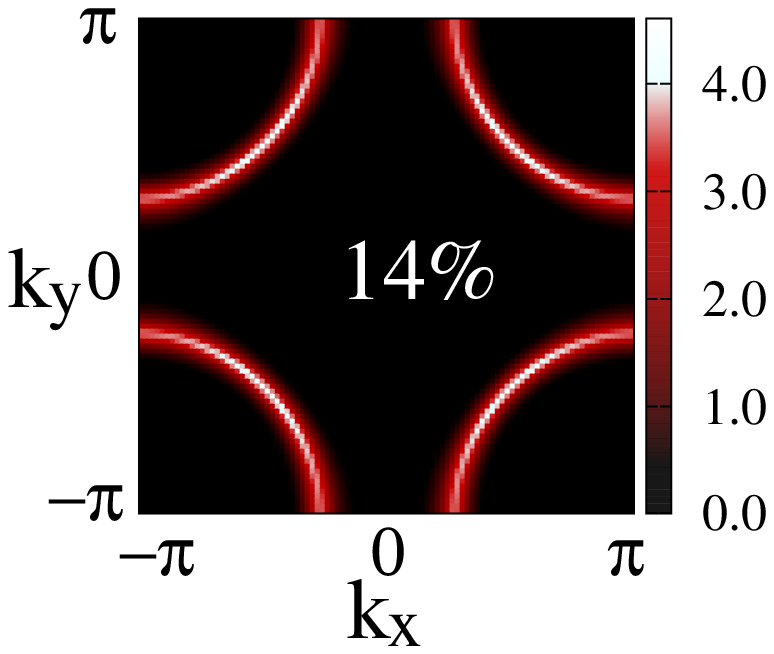}
  \includegraphics[width=4.25cm,clip=true]{fig4_4.eps}
 \caption{(Color online) Intensity maps of the spectral function $A(\vk,0)$
 for different doping levels.
 Lower-right panel: normalized intensity $A(\phi,0) / A(0,0)$ along the Fermi
 surface ($\phi = 0$ is the node, $\phi = \pm 45$ the antinode).
 The nodal intensity $A(0,0)$ is 0.045 for $\delta$=6\%,
 1.66 for $\delta$=10\% and 4.61 for $\delta$=14\%. $\beta = 200$.}
  \label{Fig_spectra_color}
\end{figure}

VB-DMFT provides a simple description of momentum differentiation
as observed in ARPES experiments.
This is illustrated by the intensity maps of the spectral function $A(\vk,0)$
(obtained with $M$-interpolation)
displayed in Fig.~\ref{Fig_spectra_color}.
At very high doping $\delta \geq 25\%$ (not shown), cluster
corrections to DMFT are negligible and
the spectral intensity is uniform along the Fermi surface.
In contrast, at lower $\delta$, momentum differentiation sets in,
revealing apparent ``Fermi arcs'' at finite temperature
with higher spectral intensity in the nodal
direction in comparison to
antinodes~\cite{damascelli_rmp_2003,parcollet_hot,civelli_breakup_prl_2005,ShenARPES_Ca_Sci2005}.
The last panel in Fig.~\ref{Fig_spectra_color} shows that the contrast
of the spectral intensity along the Fermi surface
has a maximum around $\delta\approx 10\%$, similarly to ARPES experiments (cf. Fig.~3B of~\cite{ShenARPES_Ca_Sci2005}).
At low doping, singlet formation induces a large real part in $\Sigma_{K}$ (cf. Fig.~\ref{Fig_ReSig_Weights})
and a large imaginary part of the self-energy
in the $(\pi,0)$ and $(\pi,\pi)$ regions,
which are responsible for this strong momentum-space differentiation.
At intermediate doping ($10\%\lesssim\delta\lesssim 20\%$),
this differentiation is reliably addressed using VB-DMFT.
At low doping ($\delta \lesssim 8\%$)
the $M$-interpolated self-energy develops singularities on
lines in momentum space, leading to lines of zeroes of the Green's
function and to the breakup of the Fermi surface
into pockets~\cite{StanescuCumulantCourt,Essler02,Dzyaloshinski03,Berthod06,Yang06}.
A better momentum resolution (larger clusters)
is necessary to obtain reliable
results in this regime.

VB-DMFT and the (non-self-consistent) two-impurity Anderson model
share common features. In both cases, at low-$\delta$, the singlet state dominates,
and the real part of the odd-orbital self-energy is large. These effects are due
to the term transferring singlet pairs from the even orbital to the
odd orbital, as can be checked by explicitly removing it from the dimer Hamiltonian.
Interestingly, strong fluctuations in the singlet pairing channel and
momentum-space differentiation appear to be related effects.
The key difference between VB-DMFT and the two-impurity model with fixed bath
is that the self-consistency leads to the opening of a gap in the odd orbital.
This gap reduces the scattering rate of the even orbital, leading to a maximum
in $\mathrm{Im}\Sigma_{+}(i0^+)$ (and also in the reconstructed
$\mathrm{Im}\Sigma_{\pi 0}(i0^+)$), which is absent in the non-self-consistent
two-impurity model.

To summarize, we have proposed in this article a valence-bond dynamical
mean-field theory (VB-DMFT) as a minimal cluster-based description of
momentum-space differentiation in doped Mott insulators.
Because of its simplicity, this theory can be investigated with
moderate numerical effort and progress in qualitative understanding
can be achieved with low-energy methods such as rotationally invariant
slave bosons.
The calculated STM and ARPES spectra are consistent with
the phenomenology of the normal state of cuprates.
The low-doping regime is dominated by singlet formation.
Mott physics is responsible for the suppression of
coherent quasiparticles at the antinodes, in qualitative
agreement with other approaches starting from the
weak/intermediate coupling viewpoint~\cite{Lauchli04}.
Within VB-DMFT, this suppression is described as an
orbital-selective transition in momentum-space.

\acknowledgments
We thank F. Lechermann, K. Haule and T. M. Rice for useful discussions and
acknowledge support from ICAM and the ANR under grants ETSF and GASCOR.
P.~S.~C. thanks CPHT and IPhT-Saclay for hospitality.
G.~K. was supported by the NSF and the Pascal Chair.

\end{document}